\documentclass[times,review,3p,12pt]{elsarticle}
\pdfoutput=1
\usepackage{graphicx}
\usepackage{url}
\usepackage{lineno}
\usepackage{amsmath,amstext,amssymb,amsthm} 
\usepackage{multirow,tabularx} 
\usepackage{booktabs}
\usepackage{overpic,subfigure} 
\usepackage{bm} 
\usepackage{wasysym}
\journal{International Journal of Heat and Mass Transfer}

\begin{document}

\begin{frontmatter}
\title{The piston effect induced by mass transfer in a binary fluid mixture near the critical point}
\author[coe,swum]{Zhan-Chao Hu}
\author[coe,swum]{Xin-Rong Zhang\corref{cor1}}
\ead{xrzhang@pku.edu.cn}
\cortext[cor1]{Corresponding author at: Department of Energy and Resources Engineering, College of Engineering, Peking University, Beijing 100871, China.}
\address[coe]{Department of Energy and Resources Engineering, College of Engineering, Peking University, Beijing 100871, China}
\address[swum]{Beijing Engineering Research Center of City Heat, Peking University, Beijing 100871, China}
\date{\today}

\begin{abstract}
Similar to pure fluids, physical properties of a binary fluid mixture also exhibit singularities close to its critical point, especially when it is dilute. The numerical and theoretical results presented in this paper identify that the piston effect, a rapid energy transfer phenomenon, can be induced by boundary mass transfer in a confined near-critical binary fluid mixture. Due to the Dufour effect, both the concentration and temperature variations are responsible for the strong expansion of the boundary layer, which provokes an acoustic wave propagating in the fluid, leading to a gradual increase of the temperature, pressure and density. The detailed analysis implies that such a mixed piston effect can be approximated as a direct superposition of their respective effects in the perspective of energy transformation.
\end{abstract}


\end{frontmatter}


\section{Introduction}
For a single-component fluid near the liquid-vapor critical point, the large compressibility and the diminishing thermal diffusivity are responsible for a rapid thermal relaxation process known as the piston effect (PE) \cite{Onuki1990,Zappoli1990,Boukari1990}. Due to buoyant convection on earth, the PE had long been ignored and was first observed in a microgravity experiment \cite{Nitsche1987}. Indeed, the heating of a confined near-critical fluid provokes a thin thermal boundary layer (BL), which expands strongly, acting like a piston, and drives a field of acoustic waves in the fluid. These acoustic waves travel back and forth at the speed of sound, causing adiabatic compression of the bulk fluid and rapid relaxations of temperature \cite{Carlès20102}. Asymptotic expansions were employed to study the relaxations of temperature and density fields \cite{zappoli1995,Bailly2000}. Meanwhile, several experimental studies were carried out to evidence the PE \cite{Guenoun1993,Zhong1995,Staub1995,Garrabos1998}, observe thermoacoustic waves \cite{Miura2006}, and investigate whether the PE can be used to perform long-distance heat transfer \cite{Beysens2010}. The generation and reflection of thermoacoustic waves were discussed thoroughly by Shen and Zhang \cite{Shen2010,Shen2011}. A recent study by Long et al. \cite{Long2016} discussed the thermoacoustic waves in binary fluid mixtures, with an emphasis on the influence of cross-diffusion effects. Related research progresses are widely described in a recent book written by Zappoli et al. \cite{zappoli2015}.

Generally speaking, most of previous studies focus on heat transfer and related effects in critical region. However, mass transfer, always being bracketed with heat transfer, in a critical region has not received sufficient attention. In fact, one of the most important engineering applications of near-critical and supercritical fluids is chemical extraction, in which the mass transfer is of fundamental significance. Previous studies have revealed that large density gradient and natural convection are often encountered when a solute dissolves into a near-critical fluid \cite{LIONG199191,sengers1994}. These facts motivated us to investigate whether there is any hidden phenomenon, among which our primary concern is the PE. 

Similar to pure fluids, the physical properties of a near-critical binary fluid mixture (NCBFM) exhibit singular behavior near the critical points. The theory developed by Griffiths and Wheeler \cite{Griffiths1970critical} offers a concise qualitative description for the abnormal behavior of thermodynamic properties. Variables such as pressure $p$, temperature $T$ and chemical potential $\mu$ are called field variables. Other variables such as density $\rho$, specific entropy $s$ and concentration $c$ are called density variables. A derivative of a density variable with respect to a field variable divergences strongly, with two field variables kept constant, and weakly, with one density and one field variables kept constant. The derivative of a density variable with respect to a density variable, with only field variables kept constant, is generally finite and well-behaved at a critical point. Therefore, the specific heat ratio ${c_p} = T{\left( {{{\partial s}}/{{\partial T}}} \right)_{p,c}}$, the isothermal compressibility ${\alpha _T} = {1}/{\rho }\times{\left( {{{\partial \rho }}/{{\partial p}}} \right)_{T,c}}$, 
and the thermal expansion coefficient ${\beta _p} =  - {1}/{\rho }\times{\left( {{{\partial \rho }}/{{\partial T}}} \right)_{p,c}}$ are weakly divergent in the asymptotic critical region, and the solutal expansion coefficient ${\kappa _c} =  - {1}/{\rho }\times{\left( {{{\partial \rho }}/{{\partial c}}} \right)_{p,T}}$ remains finite at the critical point. However, Sengers \cite{sengers1994} pointed out that exceptions occur in dilute binary mixtures, because they are in between of pure fluids and mixtures. The divergences of ${c_p}$, $\alpha_T$ and $\beta_p$ exhibit a crossover, changing from strong divergences to weak divergences as a critical point is approached. Besides, under the infinite-dilution condition, the partial molar volume of the solute diverges at the solvent's critical point which implies that $\kappa_c$ is also large in a near-critical dilute mixture \cite{LIONG199191}. As for transport properties, according to Luettmer-Strathmann \cite{luettmer2002} and Yang et al. \cite{Yang2000}, the diffusion coefficient $ D $ vanishes at a critical point, the viscosity $\eta$ diverges weakly, and the thermal conductivity $\lambda$ remains finite.

In this paper, we report interesting findings on the PE induced by mass transfer. We study the responses of a dilute NCBFM to boundary concentration perturbations by numerical and theoretical modeling. The hydrodynamic model, along with its solutions under different concentration perturbations on acoustic time scale are presented. Moreover, a thermodynamic theory is proposed to explain the phenomenon. 

\section{Physical and  mathematical modeling}
\subsection{Physical model and governing equations}
\begin{figure}[htb]
	\centering
	\includegraphics[width=3.4in]{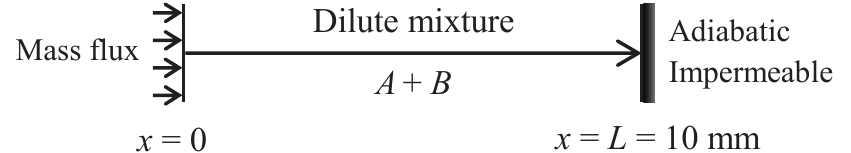}%
	\caption{\label{model}One-dimensional physical model.}
\end{figure}
We consider a dilute NCBFM of species $A$ (the minority species) and $B$ confined between two infinite solid plates that are spaced by a distance $L=10~\mathrm{mm}$ (Fig. \ref{model}). The one-dimensional model is justified since gravity is ignored (no buoyant convection and density stratification). The fluid is initially motionless at the critical density $\rho_c$ and at thermodynamic equilibrium slightly larger than its critical temperature $T_c$ (subscript $c$ denotes the critical value). A mass flux of species $ A $ happens at the left boundary, while the other side is adiabatic and impermeable. The boundary condition employed in this study is often encountered in chemical extraction processes, where species $ A $ as the solute dissolves into near-critical/supercritical solvent $B$ from solid substrate. The resulting mixtures are often dilute since the solubilities are usually small.

The Soret effect (SE) and Dufour effect (DE) should be included, since the thermal diffusion factor $k_T$, describing the SE and DE, diverges for a NCBFM \cite{luettmer2002}. The energy flux $ q $ and mass flux $ i $  are thus expressed as \cite{landau2013} (along the $ x $ direction)
\begin{eqnarray}
\label{q} q &=& \frac{{{k_T}}}{{{c_s}}}i + \overline H i - \lambda T_x ,\\
\label{i} i &=&  - \rho D{c_x} - \frac{{\rho {k_T}D}}{T}{T_x},
\end{eqnarray}
where $c$ is the concentration (mass fraction) of species $A$. The subscript $x$ represents derivative versus space. The quantity $\overline H$ is related to the partial molar enthalpy of the two species $\overline H_A$ and $\overline H_B$ by $\overline H = \overline H_A/M_A - \overline H_B/M_B$, where $M$ is the molar mass. The quantity ${c_s} = {\left( {{{\partial c}}/{{\partial \mu }}} \right)_{p,T}}$ is the concentration susceptibility ($\mu$ is related to the chemical potentials of the two species $\mu_A$ and $\mu_B$ by $\mu = \mu_A/M_A - \mu_B/M_B$). The first and second terms on the right-hand side (RHS) of Eq. (\ref{q}) are energy flux due to DE and interdiffusion of species, respectively. The second term on the RHS of Eq. (\ref{i}) is mass flux due to SE. Employing the above expressions for $q$ and $i$, the conservation equations \cite{bird2007} with constant physical properties, together with a linearized equation of state (EOS), are
\begin{eqnarray}
\label{continuty}{\rho _t} + {(\rho u)_x} &=& 0, \\
{(\rho u)_t} + {(\rho {u^2})_x} &=&  - {p_x} + \frac{{4\eta }}{3}{u_{xx}}, \\
{(\rho c)_t} + {(\rho uc)_x} &=& - {i_x}, \\
\label {T}{c_p}\left[ {{{(\rho T)}_t} + {{(\rho uT)}_x}} \right] &=& \lambda {T_{xx}}+T{\beta _p}({p_t} + u{p_x}) 
- \frac{{{k_T}}}{{{c_s}}}{i_x} + \frac{{4\eta }}{3}u_x^2, \\
\label{state}\delta \rho/\rho&=& \alpha _T\delta p - {\beta _p}\delta T - {\kappa _c}\delta c,
\end{eqnarray}
where $u$ is the velocity. The subscript $t$ represents derivative versus time. Note that in order to get concise equations, the energy flux $q$ has been fully substituted, while the mass flux $i$ remains.

\subsection{Modeling of physical properties}
The Peng-Robinson (PR) EOS together with the van der Waals mixing rule is used to obtain the thermodynamic properties of the NCBFM. In this study, we only consider the case with $ k_T,\kappa_c>0 $. The mixture of $\mathrm{C_2H_6}$ (species $A$) and $\mathrm{CO_2}$ (species $B$) with $c = 0.005 $ is chosen as a reference system because the reliable modeling by PR EOS and experimental data on critical parameters can be found in literature \cite{Sengers2007}. The reader is referred to our previous work \cite{hu_zhang_2018} and a detailed instruction \cite{pratt2001thermodynamic} for calculations of $c_p$,$\alpha$, $\beta$ and $\kappa$ based on the framework of the PR EOS and van der Waals mixing rule. The chemical potentials of components are calculated by
\begin{equation}
{\mu _A} = {{\bar \mu }_A} + RT\ln \frac{{{f_A}}}{{{{\bar f}_A}}},~{\mu _B} = {{\bar \mu }_B} + RT\ln \frac{{{f_B}}}{{{{\bar f}_B}}},
\end{equation}
where $f$ is the fugacity, $R ={\rm 8.314 ~J \,mol^{-1} K^{-1}}$ is the universal gas constant  and the overbar indicats a property at reference state ($\bar T=T$, $\bar p = \rm 100~kPa$). $f_A$ and $f_B$ are calculated directly from PR EOS and van der Waals mixing rule \cite{SZARAWARA19891489}. The reference properties are accessed from the NIST database \cite{lemmon2002}.


For transport properties, we suppose that $\eta$ and $\lambda$ of the mixture are those of pure $\mathrm{CO_2}$ at the same temperature and density (obtained from the NIST database \cite{lemmon2002}), which is a valid assumption for a dilute mixture ($c < 0.01 $).  As for $ D $ and $ k_T $, since very few experimental data is available, theoretical models should be employed. The diffusion coefficient can be expressed as \cite{LeveltSengers1993}
\begin{equation}\label{D}
D = D_s+D_b = \frac{{{k_B}T}}{{6\pi \eta \xi }} + \frac{{{\alpha _b}}}{\rho }\left( {\frac{{\partial c}}{{\partial \mu }}} \right)_{p,T}^{ - 1},
\end{equation}
where $k_B$ is the Boltzmann's constant, $\xi$ is the correlation length, $\alpha_b$ is the background part of the Onsager kinetic coefficient $\alpha$.
The diffusion coefficient $D$ in Eq. (\ref{D}) consists of a singular part $D_s$ and a background part $D_b$. Asymptotically close to the critical point, both $D_s$ and $D_b$ tend to zero due to the strong divergences of $\xi$ and the concentration susceptibility, respectively. According to Luettmer-Strathamann \cite{luettmer2002}, asymptotically close to the critical point, the following power laws hold:
\begin{eqnarray}
{D_s}  &= & \frac{{{k_B}T}}{{6\pi \eta \xi }} \sim {\xi ^{ - 1 - \phi }} = {\varepsilon ^{(1 + \phi )\nu }} \simeq {\varepsilon ^{0.67}}, \nonumber \\
{D_b} & = &\frac{{{\alpha _b}}}{\rho }\left( {\frac{{\partial c}}{{\partial \mu }}} \right)_{p,T}^{ - 1} \sim {\xi ^{-\gamma /\nu }} = {\varepsilon ^{ \gamma }} \simeq {\varepsilon ^{1.24}}, \\
D &= & D_s+D_b \sim{\varepsilon ^{0.67}}, \nonumber
\end{eqnarray}
where the universal critical exponents $ \phi $, $\nu$ and $\gamma$ have the values $\phi\simeq0.063$, $\nu\simeq0.63$ and $\gamma\simeq1.24$, and $ \varepsilon=(T-T_c)/T_c $ is the reduced temperature. The asymptotic analysis also points out the thermal diffusion factor behaves like \cite{luettmer2002}
\begin{equation}
k_T \sim \xi^{1+\phi}=\varepsilon^{-{(1+\phi)\nu}}\simeq\varepsilon^{-{0.67}}. 
\end{equation}
Therefore, in this study, we assume $D$ and $k_T$ obey the following equations:
\begin{equation}\label{powerlaw}
D=D_0{\varepsilon ^{0.67}},~~{k_T} ={k_T}_0{\varepsilon ^{ - 0.67}}.
\end{equation} 
In order to determine $D_0$ and ${k_T}_0$ in Eq. (\ref{powerlaw}), we assume critical anomalies are not noticeable when $ \varepsilon=0.1 $. Furthermore, we assume $ D|_{\varepsilon=0.1} $ can be predicted by the model developed by Vaz et al. \cite{vaz2014}, and $ k_T|_{\varepsilon=0.05}=0.05 $ (a negligible value \cite{landau2013} since the mixture is dilute). Consequently, $D_0$ and $k_{T0}$ are determined.


\subsection{Initial and boundary conditions}
Initially, the fluid is in thermodynamic equilibrium with $\varepsilon=0.003$ and $\rho=\rho_c$. The right boundary is adiabatic ($q=0$) and impermeable ($i=0$), leading to $T_x|_{x=L}=c_x|_{x=L}=u|_{x=L}=0$. For left boundary, we assume no heat conduction (namely $ T_x|_{x=0}=0 $) and two kinds of concentration perturbations are considered
\begin{equation} 
c|_{x=0} = 0.00505\ ~~ \textrm{and} ~~-c_x|_{x=0} = 500 ~\mathrm{m^{-1}} ,
\end{equation}
namely imposing a concentration step and a constant concentration gradient, respectively. The velocity at left boundary is 
\begin{equation} 
u|_{x=0} = \frac{1}{\rho(1-c)}i.
\end{equation}

We list in Table \ref{properties} the physical properties used in calculations, which are treated as constant since concentration perturbations are small.
\begin{table}[htb]
	\caption{\label{properties}The thermodynamic and transport properties of the mixture of 0.005 $ \mathrm{C_2H_6} $ and 0.995 $\mathrm{CO_2}$ at $\varepsilon=0.003$.}
	\centering
	\scriptsize
	\begin{tabular}{ccccccc}
		\toprule
		$\varepsilon$&$T_c$&$\rho_c$&$c_v$&$c_p\times10^{-4}$&$\alpha_T\times10^{6}$&$\beta_p$\\
		-&K&$\rm kg\,m^{-3}$&$\rm J\,kg^{-1}\,K^{-1}$&$\rm J\,kg^{-1}\,K^{-1}$&$\rm Pa^{-1}$&$\rm K^{-1}$\\
		\midrule
		0.003&303.73&464.89&871.05&3.95&1.60&0.31\\
		\bottomrule
		$\kappa_c$&$ c_s \times10^{8}$ & $\overline H\times10^{6}$ &$\eta\times10^{5}$&$\lambda$&$D\times10^{9}$&$k_T$\\
		-& $\rm J^{-1}\,kg$ & $ \mathrm{J~kg^{-1}} $ &$\rm Pa\cdot s$&$\rm W\,m^{-1}\, K^{-1}$&$\rm m^2\,s^{-1}$&$-$\\
		\midrule
		17.56&6.12& 2.26 &3.29&0.18&3.50&0.524\\
		\bottomrule
	\end{tabular}
\end{table}

\subsection{Numerical method}
To study the PE on acoustic time scale $ t_a = L/v_a = 40.5 ~\mathrm{\mu s} $ ($ v_a =\sqrt{c_p/c_v\times(\partial p/\partial\rho)_{T,c}}=$ 247.13 m/s is the sound speed, where $c_v$ is the specific heat at constant volume), equations (\ref{i})-(\ref{state}) with initial and boundary conditions were solved by \texttt{SIMPLE} algorithm after finite volume discretization implemented based on OpenFOAM \cite{openfoam}, an open source C++ library for computational fluid dynamics. Convective terms are discretized using a TVD (total variation diminishing) scheme with OpenFOAM's \texttt{limitedLinear} limiter. Transient terms are discretized with a first-order \texttt{Euler} scheme. The mesh, including 1080 points, was refined near the boundaries so as to accurately represent thin BLs. A time step $\Delta t = 0.01 ~\mathrm{\mu s}$ was chosen to assure proper numerical convergence of the solutions.

\subsection{Validation of the numerical code}
In 2006, Miura et al. \cite{Miura2006} observed the acoustic waves experimentally using an ultra-sensitive interferometer. Continuous heating of $\rm 1.83~ kW/m^2$ is applied during 0.2 ms to a cell filled with near-critical $\rm CO_2$. They measured density changes on a timescale of 1 $\rm \mu s$.

\begin{figure}[htb]
	\centering
	\includegraphics[width=3in]{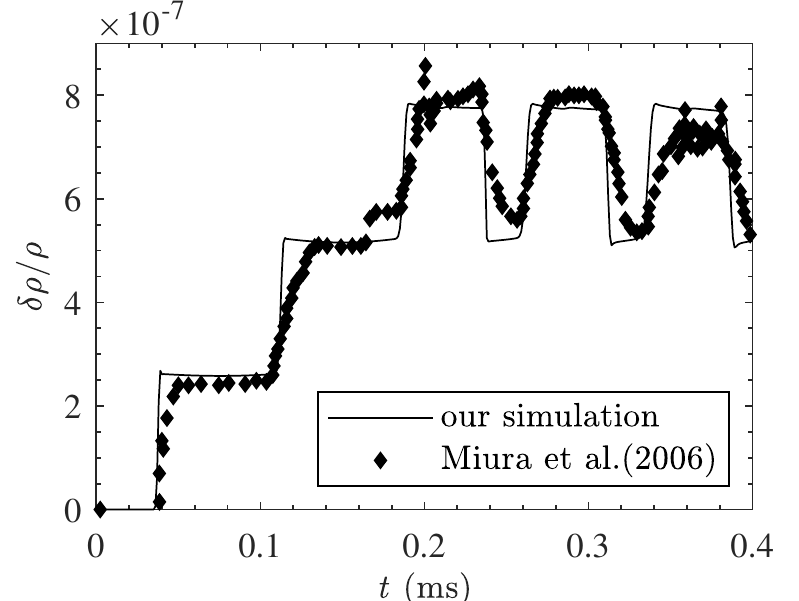}%
		\caption{\label{vali} The normalized density change at the cell center versus time. The set-up of the numerical model mimics the experiment of Miura et al. \cite{Miura2006}, serving as the validation of our numerical code.}
\end{figure}

To validate our code, a simulation has been conducted in a 1D configuration and with the same initial and boundary conditions as in the experiments. The comparisons are presented in Fig. \ref{vali}, where generally fair agreements are noticed, with overestimation in the late stage of propagation. The overestimation can be interpreted as the neglect of the damping effects of the lateral walls. 

\section{Numerical results}
The primary aim of this study is to exhibit the PE induced by mass transfer (or concentration). However, the third term on the RHS of Eq. (\ref{T}) indicates the PE is actually reinforced by the Dufour heat flux. Thus we also performed calculations with $k_T=0$ to eliminate the DE, and comparisons between two conditions are made to better explain the mechanism. Note that as indicated by Eq. (\ref{T}) when assuming a constant $ \overline H $, the interdiffusion between two species (i.e. $ \overline H i $ in Eq. (\ref{q})) actually has no influence on temperature field (absent in Eq. (\ref{T})). 

\begin{figure}[htbp]
	\centering
	\includegraphics[width=3.4in]{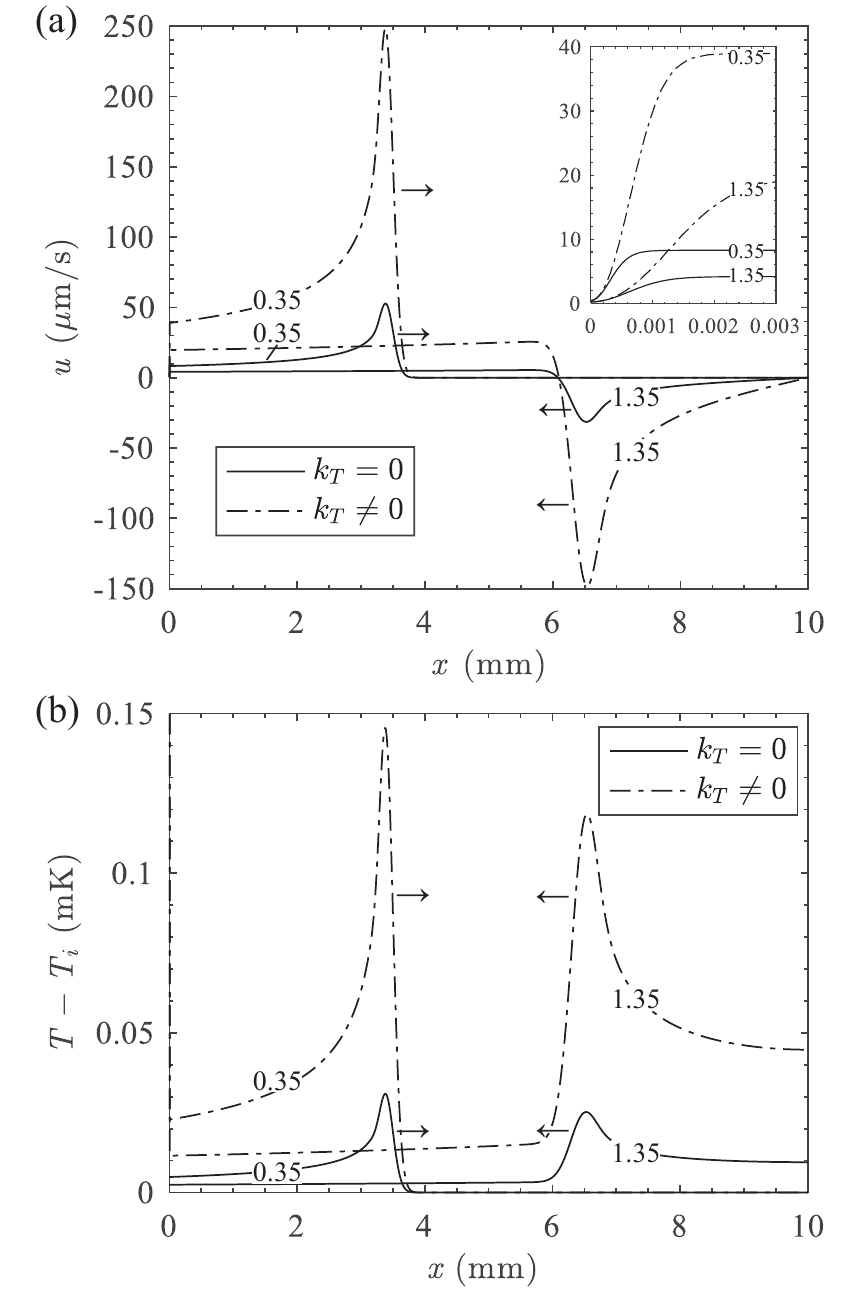}
	\caption{\label{uT} (a) Velocity and (b) temperature profiles at initial acoustic times for $ c|_{x=0} = 0.00505 $ case. $ T_i $ is the initial temperature. The inset in the upper-right corner of (a) is the enlarged left boundary region. The numbers denote the ratio $t/t_a$, with $t_a=40.5~\mathrm{\mu s}$, and arrows indicate the direction of the wave propagation. The emission and reflection of acoustic waves are clearly seen.}
\end{figure}
\begin{figure}[htbp]
	\centering
	\includegraphics[width=3.4in]{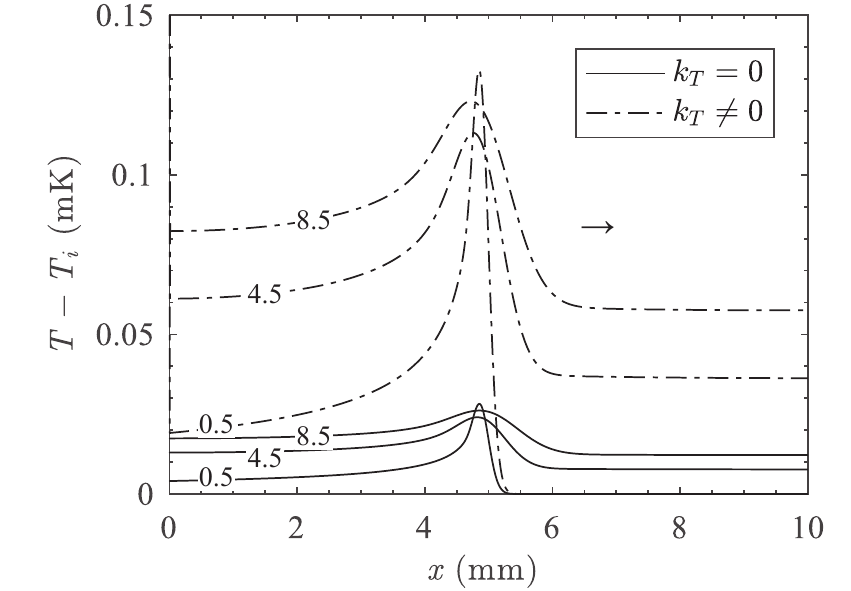}
	\caption{\label{Tlong} Temperature profiles at different acoustic times for $ c|_{x=0} = 0.00505 $ case. The numbers denote the ratio $t/t_a$ and the arrow indicates the direction of the wave propagation. The acoustic heating is clearly seen.}
\end{figure}
First consider $ c|_{x=0} = 0.00505~ $ and $k_T=0$ case, namely that a sudden concentration step of 0.00005  occurs at the left boundary and the expansion of the BL is solely caused by mass transfer. Figure \ref{uT} presents the velocity and temperature profiles at initial acoustic times (solid lines). Near the left wall, owing to the vanishing $D$ and the diverging $\kappa_c$, a thin BL expands, manifested by a steep positive velocity gradient (see the inset in Fig. \ref{uT}(a)), provoking an acoustic wave traveling in the fluid. In this process, the internal energy of the BL is transformed into the kinetic energy of the wave. As shown in Fig. \ref{uT}(a), the wave consists of two parts: a steep head in which the velocity gradient is negative (corresponding to a compression region) and a gentle tail in which the velocity gradient is positive (corresponding to an expansion region). Consequently, the $ p $, $ \rho $ and $ T $ at a fixed point in the bulk fluid first experience a sudden increase and then decrease gradually in each acoustic time. Since $ p $, $ \rho $ and $ T $ have the same behavior, only temperature profiles at different time are presented in Fig. \ref{uT}(b), whose shape is identical to that of velocity. The reflection of the acoustic wave happens when it reaches the boundaries. As the wave propagates, its kinetic energy is transformed into internal energy in the bulk fluid, making $ p $, $ \rho $ and $ T $ in the bulk fluid increase. Such an acoustic heating is clearly shown in Fig. \ref{Tlong}, which plots the temperature profiles at longer acoustic times.

When $ k_T \ne 0$ ($ k_T > 0$ in current case), a heat flux proportional to diffusion flux (given by $ k_T/c_s\cdot i $) on account of the DE is applied at the left boundary. A mixed BL then forms along left wall with both concentration and temperature gradients. However, the right edge of the BL is actually defined by temperature profile since the thermal diffusivity $D_T=\lambda/(\rho c_p)=9.94\times10^{-8} \mathrm{m^2/s}$ is larger than $D$ (the influence of SE on mass diffusion is negligible by comparison). The expansion of the BL is thus reinforced. It is observed that a stronger acoustic wave, with a similar form, propagates in the fluid (see dash-dotted lines in Fig. \ref{uT}). Such a mixed PE transfers more energy out of the BL, making the $p$, $\rho$, $T$ in the bulk fluid grow even faster (see Fig. \ref{Tlong}). 

In general, wave form is closely related to mechanical disturbance, which is imposed by the BL (the piston). The mass flux $ i $ dominates the behavior of the BL by controlling the concentration and temperature variations in it (Dufour heat flux is proportional to $ i $). The sudden increase of $c$ at left boundary induces a large $i$ (also a large Dufour heat flux) in a very short time, leading to a sudden expansion of the BL. Then the mass flux decreases gradually due to the establishment of BL, resulting in a deceleration in expansion. In other words, the BL is equivalent to a piston, which first moves rightward suddenly (causing a strong compression head in the wave) and then decelerates gradually (causing a gentle expansion tail).

\begin{figure}[htbp]
	\centering
	\includegraphics[width=3.4in]{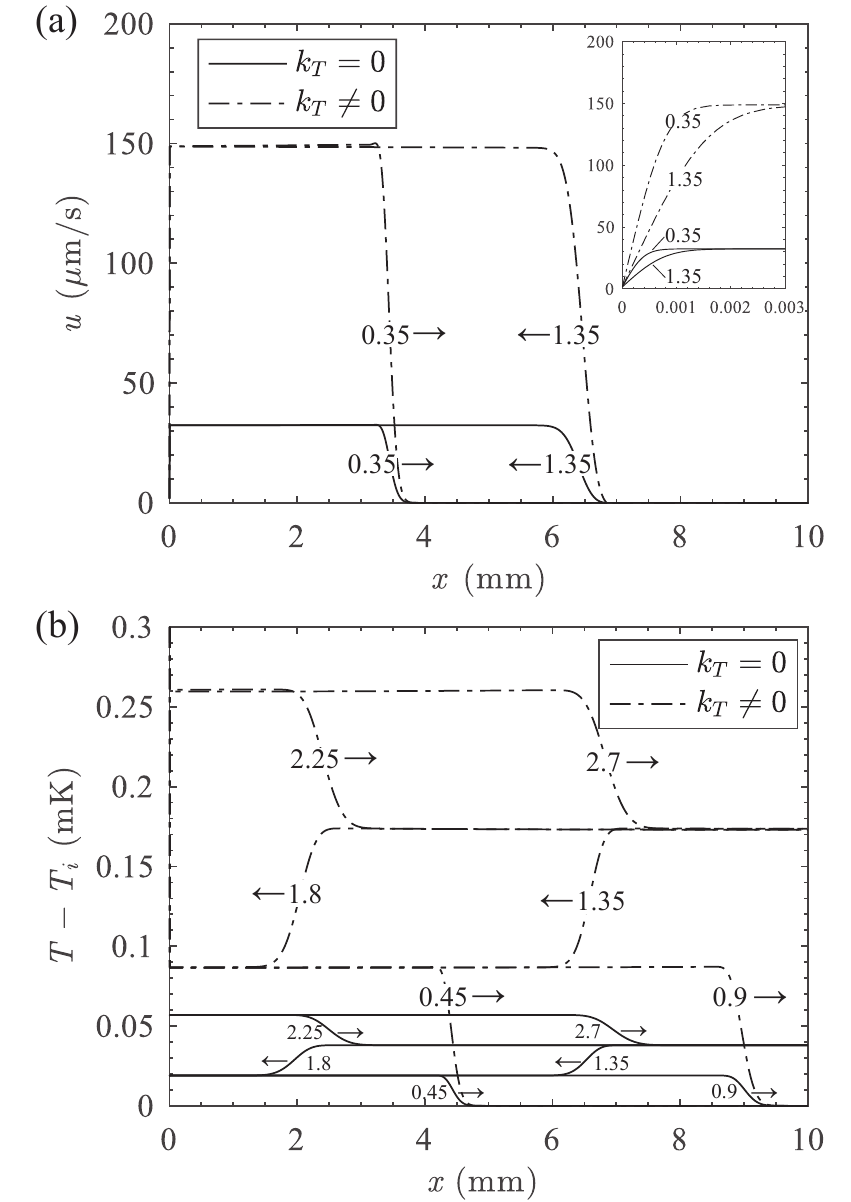}
	\caption{\label{UTflux} (a) Velocity and (b) temperature profiles at different acoustic times for $ -c_x|_{x=0} = 500 ~ \mathrm{m^{-1}} $ case (see Fig. \ref{uT} caption.)}
\end{figure}
Now consider $ -c_x|_{x=0} = 500 ~ \mathrm{m^{-1}} $ case, in which the mass flux and Dufour heat flux imposed at the left boundary are nearly constant. According to previous reasoning, one can expect a steep compression wave traveling in the fluid due to the BL expands at an almost constant speed. Presented in Fig. \ref{UTflux} is the velocity and temperature profiles at different acoustic times. The expanding BL is shown in the inset of Fig. \ref{UTflux}(a), where the velocity gradient is positive. In agreement with above prediction, it is observed that an abrupt compression wave (corresponding to a negative velocity gradient) propagates in the fluid, which reflects back at the right boundary (see Fig. \ref{UTflux}(a)). The acoustic heating is clearly shown in Fig. \ref{UTflux}(b), the compression wave makes a traveling increase in the bulk temperature. When $k_T\ne0$, the reinforced PE is evidenced by the stronger compression wave and acoustic heating.

\section{Theory}
\subsection{Expressions for amplitudes}\label{appendix}
We give a thermodynamic theory to explain the phenomenon for constant concentration gradient case, obtaining the expressions of amplitudes of velocity, density, pressure, temperature, denoted as $u'$, $\rho'$, $p'$, $T'$, respectively \cite{Onuki1990,Miura2006}. Imagining the BL as a moving piston with a constant speed, $u'$ should equal the piston's velocity, namely the volume expansion rate per unit area of the BL. According to the linearized equation of state (see Eq. (\ref{state})), expansion of the boundary layer is the result of concentration increase and temperature increase (due to Dufour heat flux). Thus the expression of $u'$ should be a superposition of the two above-mentioned factors. 

From a thermodynamic point of view, the relaxation of concentration field $c$ is dominated by diffusion. Further ignoring the Soret effect, the governing equation, along with the initial and boundary conditions, is
\begin{eqnarray}
&& \frac{{{\rm{d}}c}}{{{\rm{d}}t}}=D\frac{{{{\rm{d}}^2}c}}{{{\rm{d}}{x^2}}}, \nonumber   \\
&& t = 0,c=c_0,             \\
&& x=0, -c_x=g_0.\nonumber
\end{eqnarray}
Because the diffusion coefficient $D$ is quite small near the critical point, the domain length $ L=10 $ mm is much larger compared to the boundary layer thickness $l_{BL}$ on acoustic time, leading to the following solution \cite{Incropera2006}
\begin{equation}\label{2}
c(x,t) - {c_0} = 2{g_0}\sqrt {\frac{{Dt}}{\pi }} \exp \left( { - \frac{{{x^2}}}{{4Dt}}} \right) - {g_0}x{\rm{erfc}}\left( {\frac{x}{{2\sqrt {Dt} }}} \right). 
\end{equation}
At the edge of concentration boundary layer $x=l_{BL}$, the following equation holds
\begin{equation}\label{3}
\frac{{c(0,t) - c(x = {l_{BL}},t)}}{{c(0,t) - {c_0}}} = 0.99.
\end{equation}
Using Eq. (\ref{2}) and defining $\omega = x/(2\sqrt{Dt})$, Eq. (\ref{3}) is transformed to
\begin{equation}\label{4}
\exp ( - {\omega ^2}) - \sqrt \pi  \omega {\rm{erfc(}}\omega)= 0.01,
\end{equation}
whose numerical solution is $ \omega \approx 1.606 $. Thus we obtain
\begin{equation}\label{5}
l_{BL}\approx3.211\sqrt{Dt}.
\end{equation}
The volume change of the boundary layer per unit area due to concentration increase is given by
\begin{equation}\label{6}
\delta V_c(t)=\int_0^{{l_{BL}}} {{\kappa _c}\left[ {c(x,t) - {c_0}} \right]} {\rm{d}}x.
\end{equation}
Inserting  Eq. (\ref{2}) and Eq. (\ref{5}) into Eq. (\ref{6}), and using $w$ to replace $x$, we obtain
\begin{equation}\label{7}
\delta V_c(t) = 4D{\kappa _c}{g_0}t\int_0^{1.606} {\left[ {\frac{1}{{\sqrt \pi  }}\exp ( - {\omega ^2}) - \omega {\rm{erfc(}}\omega {\rm{)}}} \right]} {\rm{ d}}\omega.
\end{equation}
The approximate value of the integral in Eq. (\ref{7}) is 0.249. Therefore, $ \delta V(t) $ is approximated by
\begin{equation}
\delta V_c(t) \approx D\kappa_cg_0t.
\end{equation}
The velocity amplitude due to concentration increase is 
\begin{equation}\label{9}
u'_c=\frac{\mathrm{d}\delta V_c(t)}{\mathrm{d}t} = D\kappa_cg_0.
\end{equation}

As for temperature increase induced expansion, we follow Onuki's approach \cite{Onuki1990}. The Dufour heat flux at left boundary is given by 
\begin{equation}
q_D = \frac{\rho k_TD}{c_s} g_0.
\end{equation}
Assuming expansion of the boundary layer is an isobaric process \cite{Onuki1990}, the volume change of the boundary layer per unit area due to temperature increase is given by 
\begin{equation}
\delta V_T(t)=\frac{q_D}{\rho l_{BL}c_p}\beta_pl_{BL}t= \frac{{{k_T}D{\beta _p}g_0}}{{{c_s}{c_p}}}t.
\end{equation}
The velocity amplitude due to temperature increase is 
\begin{equation}\label{12}
u'_T=\frac{\mathrm{d}\delta V_T(t)}{\mathrm{d}t} = \frac{{{k_T}D{\beta _p}g_0}}{{{c_s}{c_p}}}.
\end{equation}

Finally, we obtain the expression for velocity amplitude of the acoustic wave
\begin{eqnarray}\label{up}
u' &=& u'_c+u'_T,\nonumber \\
&=& D\kappa_cg_0+\frac{{{k_T}D{\beta _p}g_0}}{{{c_s}{c_p}}}, \\
&=& -D{\kappa_c}{c_x|_{x=0}} - \frac{{{k_T}D{\beta _p}{c_x|_{x=0}}}}{{{c_s}{c_p}}}.\nonumber 
\end{eqnarray}
The superposition is justified since their separate contributions in the expansion of the BL. Note that for the sake of simplicity, the SE is ignored in Eq. (\ref{up}), which results in a slight undervaluation.

The mass conservation gives $\rho'$ (ignore small quantity of the second order)
\begin{equation}
\rho ' = \frac{\rho }{{{v_a}}}u'.
\end{equation}
The isentropic nature of the acoustic wave gives the expressions for $p'$ and $T'$
\begin{eqnarray}
p' &=& {\left( {\frac{{\partial p}}{{\partial \rho }}} \right)_s}\rho ' = \rho {v_a}u',\\
T' &=& {\left( {\frac{{\partial T}}{{\partial p}}} \right)_s}p' = \frac{{{\beta _p}T{v_a}}}{{{c_p}}}u'.
\end{eqnarray}

\subsection{Energy and temperature efficiencies}
The energy efficiency $ \zeta_E $ of PE is defined as the ratio of energy transferred into bulk fluid $E_{bulk}(t)$ and energy sent into the fluid $E_{in}(t)$, where\[E_{bulk}(t)=\int_{bulk} {p{\delta \rho }/\rho } {\rm{d}}x = pu't,\] \[{E_{in}}(t)=\rho D\left( {{{k_T}}}/{{{c_s}} + \overline H } \right){c_x}t,\] namely
\begin{equation}\label{effE}
\zeta_E  = \frac{{{E_{bulk}}(t)}}{{{E_{in}}(t)}} = \frac{p}{{\rho \left( {{{{k_T}}}/{{{c_s}}} + \overline H } \right)}}\left( {\frac{{{k_T}{\beta _p}}}{{{c_s}{c_p}}} + {\kappa_c}} \right).
\end{equation}
The temperature efficiency $ \zeta_T $ of PE is the ratio of actual temperature increase $\Delta T(t)=\overline T(t)-T_i=T't/t_a$ ($ \overline T $ is the average temperature) and the ideal temperature increase $\Delta T_{ideal}(t)={E_{in}}(t)/(\rho Lc_v)$, leading to
\begin{equation}\label{effT}
{\zeta _T} = \frac{{\Delta T(t)}}{{\Delta {T_{ideal}}(t)}} = \frac{{{\beta _p}T}}{{\rho {\alpha _T}\left( {{k_T}/{c_s} + \overline H} \right)}}\left( {\frac{{{k_T}{\beta _p}}}{{{c_s}{c_p}}} + {\kappa _c}} \right).
\end{equation}
We compare theoretical values and those obtained from numerical results in Table \ref{theory}. Above relations are validated by the good agreements between numerical and theoretical values. $ \zeta_E $ and $ \zeta_T $ in current problem are almost the same with those of thermal PE, whose $ \zeta_E=p/T \cdot {\left( {\partial T/\partial p} \right)_{s,c}} $ \cite{Miura2006} is equal to 0.1260 and $ \zeta_T $ is very close to 1. The discrepancy between $ \zeta_E $ and $ \zeta_T $ can be explained by the fact that the internal energy of a near-critical fluid mainly depends on density \cite{Garrabos1998}, but the PE makes little contribution to the relaxation of density field. 
\begin{table}
\caption{\label{theory} Amplitudes of the acoustic wave for $ -c_x|_{x=0} = 500 ~ \mathrm{m^{-1}} $ case obtained from numerical results (columns denoted as Simulation) and calculated from theoretical expressions Eqs. (\ref{up})-(\ref{effT})
 (columns denoted as Theory).}
\centering
\begin{tabular}{cccccc}
	\toprule
	 &&\multicolumn{2}{c}{$k_T=0$}&\multicolumn{2}{c}{$k_T\ne0$}\\
	 &unit&Simulation&Theory&Simulation&Theory\\ 
	 \midrule
	 $u'$&$ \mathrm{\mu m/s} $&32.43&30.70&148.88&147.25\\
	 $\rho'$&$ \mathrm{g/m^3} $&0.0607&0.0578&0.278&0.277\\
	 $p'$&Pa&3.73&3.53&17.08&16.92\\
	 $T'$&mK&0.0190&0.0180&0.0870&0.0861\\
	 $\zeta_E$&-&$ \sim $ 0.1331&0.1264&$ \sim $0.1262&0.1261\\
	 $\zeta_T$&-& $ \sim $ 0.9838 &0.9814& $ \sim $0.9775 &0.9787\\
	 \bottomrule
\end{tabular}
\end{table}

To further understand the coupling between the PEs induced by concentration and temperature, we decompose Eq. (\ref{effE}) as follows:
\begin{equation}\label{decompose}
\zeta_E=\frac{{{k_T}/{c_s}}}{{{k_T}/{c_s} + \overline H}}{{\zeta _{E}}}_T{\rm{ + }}\frac{{\overline H}}{{{k_T}/{c_s} + \overline H}}{\zeta _{{E}}}_c,
\end{equation}
where
\begin{eqnarray}
\label{eq:effE_1}{\zeta _E}_T &=& \frac{p}{T}{\left( {\frac{{\partial T}}{{\partial p}}} \right)_{s,c}}, \\
\label{eq:effE_2}{\zeta _E}_c &=& \left[ \frac{{\rho \mu }}{{p{\kappa _c}}} + \frac{T}{p}{\left( {\frac{{\partial p}}{{\partial T}}} \right)_{s,c}} \right] ^{-1},
\end{eqnarray}
where $ {\zeta _E}_T $ is the energy efficiency of the PE solely driven by the Dufour heat flux, and $ {\zeta _E}_c $ is the energy efficiency of the PE solely driven by concentration variations. The coefficients in Eq. (\ref{decompose}) represent the ratios of their respective driving energy to total energy. The decomposition of $\zeta_E$ implies that in the perspective of energy transformation, the mixed PE is a direct superposition of their respective effects. In fact, this conclusion is based on the approximation that properties of the fluid mixture are constant. Actually, the governing equations suggest that concentration and temperature are coupled nonlinearly through the SE and the DE, which influence properties of the fluid by changing the distributions of concentration and temperature fields, and further the total amount of energy entering the fluid. In this study, these variations in properties are neglected since the perturbations applied at the left boundary are small. Besides, Eq. (\ref{eq:effE_2}) indicates that $ {\zeta _E}_c $ is greater than $ {\zeta _E}_T $ when $\mu/\kappa_c <0$. Even though the difference between $ {\zeta _E}_c $ (equals 0.1264) and $ {\zeta _E}_T $ (equals 0.1260) is imperceptible for $ \mathrm{C_2H_6-CO_2} $ mixture, the above conclusion implies a potential in improving the energy efficiency by carefully choosing the mixture. 

\section{Conclusions}
The numerical results and theory presented in this paper clearly exhibit the PE, a rapid energy transport mechanism, can be induced by mass transfer in a confined NCBFM. In current problem, the diffusion of one species, together with the Dufour heat flux, makes the BL expand and drives an acoustic wave propagating in the fluid, making the temperature, pressure and density increase over several acoustic times. By this means, energy is transferred from the BL to bulk fluid with an efficiency $\zeta_E$ of about 0.126. For the sake of clarity, we call the PE induced by mass transfer (or concentration) as \textit{solutal piston effect} (SPE). It should be kept in mind that SPE is always coupled with thermal PE owing to the DE. The resulting mixed PE can be approximated as a direct superposition of their respective effects in the perspective of energy transformation.

This paper extends the concept of the PE and gives new insights into fluid behavior in the critical region. In the future, the SPE for different fluid systems is worth investigating, in which the signs of $\kappa_c$ and $k_T$ may be different and the BL would expand or contract according to the specific conditions. Moreover, it is also necessary to explore the interplay between the SPE with natural convection. 


\section*{Acknowledgement}
This work is supported by the Natural Science Foundation of China (Grant No. 51776002). 


\begin{thebibliography}{10}
	
	\bibitem{Onuki1990}
	A.~Onuki, H.~Hao, R.~A. Ferrell, Fast adiabatic equilibration in a
	single-component fluid near the liquid-vapor critical point, Phys. Rev. A 41
	(1990) 2256--2259.
	
	\bibitem{Zappoli1990}
	B.~Zappoli, D.~Bailly, Y.~Garrabos, B.~Le~Neindre, P.~Guenoun, D.~Beysens,
	Anomalous heat transport by the piston effect in supercritical fluids under
	zero gravity, Phys. Rev. A 41 (1990) 2264--2267.
	
	\bibitem{Boukari1990}
	H.~Boukari, J.~N. Shaumeyer, M.~E. Briggs, R.~W. Gammon, Critical speeding up
	in pure fluids, Phys. Rev. A 41 (1990) 2260--2263.
	
	\bibitem{Nitsche1987}
	K.~Nitsche, J.~Straub, Critical '{HUMP}' of $c_v$ under microgravity results
	from the {D1}-spacelab experiment '{WAERMEKAPAZITAET}', European Space Agency
	(1987) 109--116.
	
	\bibitem{Carlès20102}
	P.~Carl\`{e}s, A brief review of the thermophysical properties of supercritical
	fluids, The Journal of Supercritical Fluids 53~(1–3) (2010) 2 -- 11.
	
	\bibitem{zappoli1995}
	B.~Zappoli, P.~Carles, The thermo-acoustic nature of the critical speeding up,
	European journal of mechanics. B, Fluids 14~(1) (1995) 41--65.
	
	\bibitem{Bailly2000}
	D.~Bailly, B.~Zappoli, Hydrodynamic theory of density relaxation in
	near-critical fluids, Phys. Rev. E 62 (2000) 2353--2368.
	
	\bibitem{Guenoun1993}
	P.~Guenoun, B.~Khalil, D.~Beysens, Y.~Garrabos, F.~Kammoun, B.~Le~Neindre,
	B.~Zappoli, Thermal cycle around the critical point of carbon dioxide under
	reduced gravity, Phys. Rev. E 47 (1993) 1531--1540.
	
	\bibitem{Zhong1995}
	F.~Zhong, H.~Meyer, Density equilibration near the liquid-vapor critical point
	of a pure fluid: Single phase {$T>{\mathit{T}}_{\mathit{c}}$}, Phys. Rev. E
	51 (1995) 3223--3241.
	
	\bibitem{Staub1995}
	J.~Straub, L.~Eicher, A.~Haupt, Dynamic temperature propagation in a pure fluid
	near its critical point observed under microgravity during the german
	spacelab mission {D-2}, Phys. Rev. E 51 (1995) 5556--5563.
	
	\bibitem{Garrabos1998}
	Y.~Garrabos, M.~Bonetti, D.~Beysens, F.~Perrot, T.~Fr\"ohlich, P.~Carl\`es,
	B.~Zappoli, Relaxation of a supercritical fluid after a heat pulse in the
	absence of gravity effects: Theory and experiments, Phys. Rev. E 57 (1998)
	5665--5681.
	
	\bibitem{Miura2006}
	Y.~Miura, S.~Yoshihara, M.~Ohnishi, K.~Honda, M.~Matsumoto, J.~Kawai,
	M.~Ishikawa, H.~Kobayashi, A.~Onuki, High-speed observation of the piston
	effect near the gas-liquid critical point, Phys. Rev. E 74 (2006) 010101.
	
	\bibitem{Beysens2010}
	D.~Beysens, D.~Chatain, V.~S. Nikolayev, J.~Ouazzani, Y.~Garrabos, Possibility
	of long-distance heat transport in weightlessness using supercritical fluids,
	Phys. Rev. E 82 (2010) 061126.
	
	\bibitem{Shen2010}
	B.~Shen, P.~Zhang, On the transition from thermoacoustic convection to
	diffusion in a near-critical fluid, International Journal of Heat and Mass
	Transfer 53~(21-22) (2010) 4832--4843.
	
	\bibitem{Shen2011}
	B.~Shen, P.~Zhang, Thermoacoustic waves along the critical isochore, Phys. Rev.
	E 83~(1) (2011) 011115.
	
	\bibitem{Long2016}
	Z.~Long, P.~Zhang, B.~Shen, Thermomechanical effects in supercritical binary
	fluids, International Journal of Heat and Mass Transfer 99 (2016) 470 -- 484.
	
	\bibitem{zappoli2015}
	B.~Zappoli, D.~Beysens, Y.~Garrabos, et~al., Heat Transfers and Related Effects
	in Supercritical Fluids, Springer, 2015.
	
	\bibitem{LIONG199191}
	K.~Liong, P.~Wells, N.~Foster, Diffusion in supercritical fluids, The Journal
	of Supercritical Fluids 4~(2) (1991) 91 -- 108.
	
	\bibitem{sengers1994}
	J.~M. H.~L. Sengers, Critical Behavior of Fluids: Concepts and Applications,
	Springer Netherlands, Dordrecht, 1994, pp. 3--38.
	
	\bibitem{Griffiths1970critical}
	R.~B. Griffiths, J.~C. Wheeler, Critical points in multicomponent systems,
	Phys. Rev. A 2 (1970) 1047--1064.
	
	\bibitem{luettmer2002}
	J.~Luettmer-Strathmann, Thermodiffusion in the critical region, Springer Berlin
	Heidelberg, Berlin, Heidelberg, 2002, pp. 24--37.
	
	\bibitem{Yang2000}
	X.-n. Yang, L.~A.~F. Coelho, M.~A. Matthews, Near-critical behavior of mutual
	diffusion coefficients for five solutes in supercritical carbon dioxide,
	Industrial \& Engineering Chemistry Research 39~(8) (2000) 3059--3068.
	
	\bibitem{landau2013}
	L.~Landau, E.~Lifshitz, Fluid Mechanics, no.~6, Elsevier Science, 2013.
	
	\bibitem{bird2007}
	R.~B. Bird, W.~E. Stewart, E.~N. Lightfoot, Transport phenomena, John Wiley \&
	Sons, 2007.
	
	\bibitem{Sengers2007}
	J.~V. Sengers, G.~X. Jin, A note on the critical locus of mixtures of carbon
	dioxide and ethane, International Journal of Thermophysics 28~(4) (2007)
	1181--1187.
	
	\bibitem{hu_zhang_2018}
	Z.-C. Hu, X.-R. Zhang, Onset of convection in a near-critical binary fluid
	mixture driven by concentration gradient, Journal of Fluid Mechanics 848
	(2018) 1098–1126.
	
	\bibitem{pratt2001thermodynamic}
	R.~Pratt, Thermodynamic properties involving derivatives: Using the
	{Peng-Robinson} equation of state, Chemical Engineering Education 35~(2)
	(2001) 112--139.
	
	\bibitem{SZARAWARA19891489}
	J.~Szarawara, A.~Gawdzik, Method of calculation of fugacity coefficient from
	cubic equations of state, Chemical Engineering Science 44~(7) (1989) 1489 --
	1494.
	
	\bibitem{lemmon2002}
	E.~W. Lemmon, M.~L. Huber, M.~O. McLinden, {NIST} reference fluid thermodynamic
	and transport properties--{REFPROP}, NIST standard reference database 23
	(2002) v7.
	
	\bibitem{LeveltSengers1993}
	J.~M.~H. Levelt~Sengers, U.~K. Deiters, U.~Klask, P.~Swidersky, G.~M.
	Schneider, Application of the taylor dispersion method in supercritical
	fluids, International Journal of Thermophysics 14~(4) (1993) 893--922.
	
	\bibitem{vaz2014}
	R.~V. Vaz, A.~L. Magalh{\~a}es, C.~M. Silva, Prediction of binary diffusion
	coefficients in supercritical {${\rm CO_2}$} with improved behavior near the
	critical point, The Journal of Supercritical Fluids 91 (2014) 24--36.
	
	\bibitem{openfoam}
	\url{https://openfoam.org}.
	
	\bibitem{Incropera2006}
	F.~P. Incropera, Fundamentals of Heat and Mass Transfer, John Wiley \& Sons,
	2006.
	
\end{thebibliography}

\end{document}